\documentclass[conference]{IEEEtran}
\IEEEoverridecommandlockouts
\usepackage{spconf}

\makeatletter
\renewcommand\section{\@startsection {section}{1}{\z@}%
{1.5ex plus 1.5ex minus 0.5ex}{0.7ex plus 1ex minus 0ex}%
{\centering\normalfont\normalsize\bfseries}}
\makeatother

\normalsize
\usepackage{amsmath,amssymb,amsthm}
\usepackage{enumerate}
\usepackage{stfloats}
\usepackage{comment}
\usepackage{subcaption}
\usepackage{siunitx}
\usepackage{mathtools}
\usepackage{accents,latexsym}
\usepackage{cite}

\usepackage[dvipsnames]{xcolor}
\definecolor{myPink}{RGB}{255,105,183}

\usepackage[T1]{fontenc}
\usepackage{graphics} 
\usepackage{epsfig} 
\usepackage[mathscr]{euscript}
\usepackage{algorithm}
\usepackage[noend]{algpseudocode}
\usepackage{bbm}
\makeatletter
\def\BState{\State\hskip-\ALG@thistlm}
\makeatother

\usepackage{tikz}
\usetikzlibrary{arrows,shapes,chains,matrix,positioning,scopes,patterns,calc,
decorations.markings,
decorations.pathmorphing,
}

\usepackage{pgfplots}
\pgfplotsset{compat=1.3}
\usepgflibrary{shapes}

\newtheorem{theorem}{Theorem}

\newtheorem{proposition}[theorem]{Proposition}

\renewcommand{\epsilon}{\varepsilon}

\newcommand{\RNum}[1]{\uppercase\expandafter{\romannumeral #1\relax}}

\newcommand{\bv}{\ensuremath{\underline{b}}}

\newcommand{\pv}{\ensuremath{\underline{p}}}

\newcommand{\vv}{\ensuremath{\underline{v}}}

\newcommand{\wv}{\ensuremath{\underline{w}}}
\newcommand{\xv}{\ensuremath{\underline{x}}}
\newcommand{\yv}{\ensuremath{\underline{y}}}
\newcommand{\zv}{\ensuremath{\underline{z}}}

\newcommand{\Ka}{\ensuremath{K_{\mathrm{a}}}}

\def\Pr{\mathrm{Pr}}

\DeclareMathAlphabet{\mcl}{OMS}{cmsy}{m}{n}

\newlength\tikzwidth
\newlength\tikzheight

\definecolor{mycolor1}{rgb}{0.63529,0.07843,0.18431}%
\definecolor{mycolor2}{rgb}{0.00000,0.44706,0.74118}%
\definecolor{mycolor3}{rgb}{0.00000,0.49804,0.00000}%
\definecolor{mycolor4}{rgb}{0.87059,0.49020,0.00000}%
\definecolor{mycolor5}{rgb}{0.00000,0.44700,0.74100}%
\definecolor{mycolor6}{rgb}{0.74902,0.00000,0.74902}%

\newif\ifproof
\prooftrue

\def\fig_path{./Figures}
\begin{document}
\title{An Enhanced Decoding Algorithm for Coded Compressed Sensing}
%
\name{\large{
Vamsi K. Amalladinne,
Jean-Francois Chamberland,
Krishna R. Narayanan}
\thanks{This material is based upon work supported, in part, by the National Science Foundation (NSF) under Grant No.~CCF-1619085 and by Qualcomm Technologies, Inc., through their University Relations Program.}}
\address{Electrical and Computer Engineering, Texas A\&M University\\College Station, TX 77843, USA}
\maketitle

\begin{abstract}
Coded compressed sensing is an algorithmic framework tailored to sparse recovery in very large dimensional spaces.
This framework is originally envisioned for the unsourced multiple access channel, 
a wireless paradigm attuned to machine-type communications.
Coded compressed sensing uses a divide-and-conquer approach to break 
the sparse recovery task into 
sub-components whose dimensions are amenable to conventional compressed sensing solvers.
The recovered fragments are then stitched together using a low complexity decoder.
This article introduces an enhanced decoding algorithm for coded compressed sensing where fragment recovery 
and the stitching process are executed in tandem, passing information between them.
This novel scheme leads to gains in performance and a significant reduction in computational complexity.
This algorithmic opportunity stems from the realization that the parity structure inherent to coded compressed sensing can be used to dynamically restrict the search space of the subsequent recovery algorithm.
\end{abstract}

\section{\uppercase{Introduction}}
\label{section:Introduction}

The emergence of machine-driven wireless communications and the Internet of Things (IoT) are poised to disrupt existing communication infrastructures.
To ready wireless systems for such a transformation, new communication models are being introduced, along with novel access schemes.
Notably, the unsourced multiple access communication (MAC) channel was proposed by Polyanskiy in~\cite{polyanskiy2017perspective} to accommodate the sporadic transmission of short packets.
Along with this new perspective, Polyanskiy also introduced an achievability bound for finite blocklength communication over the unsourced MAC.
This bound is derived in the absence of computational complexity constraints and has served as a benchmark for pragmatic schemes~\cite{ordentlich2017low, vem2019user, pradhan2019sparseidma, marshakov2019polar}.

In~\cite{amalladinne2018couple}, we proposed a complexity reduction technique for the unsourced MAC based on splitting data into fragments.
This framework, called coded compressed sensing (CCS), leverages the strong connection between the unsourced MAC and compressed sensing (CS) in high dimensions.
The gist of the approach is to break a CS problem with exceedingly large dimensionality into manageable sub-components.
For the unsourced MAC, this translates into sending sequences of fragments, one per slot, rather than the entire payload.
A commodity CS solver can then be applied to every slot.
Yet, the output of the recovery process yields a collection of unordered list of message fragments, rather than a collection of messages.
That is, fragments coming from a same message must be pieced together.
To enable this process, redundancy in the form of parity bits is added to every fragment.
The resulting message structure is then employed by a tree decoder to stitch message together.
The algorithm is described and analyzed thoroughly in~\cite{amalladinne2019coded}.

Despite its recent introduction, the CCS framework has drawn attention.
In~\cite{calderbank2018chirrup}, Calderbank and Thompson combine the CCS framework with a low-complexity CS construction based on second order Reed-Muller codes~\cite{howard2008fast} to create an ultra-low complexity CS scheme.
Also, recent work by Fengler, Jung, and Caire~\cite{Giuseppe} draws a close connection between the sparse structure created by CCS and sparse regression codes (SPARCs)~\cite{joseph2013fast,CIT-092,rush2017capacity}.
Therein, they leverage the CCS data structure, but pair it with a dense CS matrix (rather than the CCS block diagonal structure) and employ approximate message passing (AMP)~\cite{bayati2011dynamics,barbier2017approximate} to decode it.

The decoding algorithm proposed for the CCS framework features two components, namely sparse recovery and fragment stitching.
In our original treatment and later contributions~\cite{amalladinne2019coded,calderbank2018chirrup,Giuseppe,fengler2019massive}, these tasks are treated separately.
Support recovery is performed first, followed by stitching through tree decoding.
However, it has become apparent that the information contained in message fragments in the form of parity bits can be integrated into the recovery process.
In particular, consistent partial paths found in the decoding tree collectively restrict the realm of possibilities for parity bits in subsequent fragments.
Based on this information, the CS matrix associated with slots can be pruned dynamically in the standard CCS framework.
Likewise, these conditions can be embedded in the graphical representation of the problem used for AMP in~\cite{Giuseppe}.
In this article, we focus on the former and present an algorithmic enhancement to the standard CCS decoder.
The result is a dynamic where fragment recovery and message stitching work synergistically, leading to complexity reduction and performance improvements.

\section{\uppercase{System Model and Background}}
\label{section:SystemModel}

The unsourced MAC model seeks to capture sporadic communications from many devices to an access point.
The motivation behind this channel can be found in~\cite{polyanskiy2017perspective}.
For the discussion at hand, it suffices to mention that this model admits a CS representation.
Specifically, in its most basic form, the unsourced MAC problem is captured by the equation
\begin{equation} \label{syseq}
\textstyle \yv= \sum_{i \in \mathbf{S}_\mathrm{a}} \xv_i + \zv
\end{equation}
where $\xv_i$ is the $n$-dimensional codeword corresponding to message~$i$, and $\zv$ denotes additive white Gaussian noise with covariance $\sigma^2 \mathbf{I}$.
The collection of $B$-bit information messages transmitted on the channel is $W = \left\{ \wv_i : i \in \mathbf{S}_\mathrm{a} \right\}$, where $|\mathbf{S}_\mathrm{a}|= \Ka$.
All the devices employ the same codebook and, as such, $\xv_i = f(\wv_i)$, irrespective of the device performing the encoding.
The decoding task is to produce a list estimate $\widehat{W}(\yv)$ for the transmitted messages $W$ with $| \widehat{W}(\yv) | \le \Ka$.
Overall performance is assessed using the per-user error probability defined by
\begin{equation}
\textstyle P_{\mathrm{e}} = \frac{1}{\Ka} \sum_{i \in \mathbf{S}_\mathrm{a}}
\Pr \left( \wv_i \notin \widehat{W}(\yv) \right) .
\end{equation}

The CS analogy for this problem is obtained by using an alternate message representation.
Suppose that we interpret $B$-bit message $\wv$ as a location in a vector of length $2^B$.
That is, this latter vector has zeros everywhere except for a one at location $\left[ \wv \right]_2$, where $[\cdot]_2$ denotes an integer expressed with a radix of 2 (binary form).
We call this latter form the message index.
To each such vector corresponds a signal $\xv = f(\wv)$.
If we build matrix $\mathbf{X} \in  \mathbb{R}^{n \times 2^B}$ where the columns are codewords $\left\{ f(\wv) \right\}$ in ascending $\wv$ order, then we can rewrite \eqref{syseq} as
\begin{equation} \label{equation:CSInterpretation}
\yv= \mathbf{X} \bv + \zv .
\end{equation}
In this characterization, $\bv$ is a $\Ka$-sparse vector that is equal to the sum of the transmitted message indices.
While \eqref{equation:CSInterpretation} assumes the form of a noisy CS problem, the sheer dimension of the problem precludes the direct application of commodity CS solvers.
The goal of CCS is to offer pragmatic encoding and decoding schemes that together achieve $P_\mathrm{e} \le \varepsilon$, where $\varepsilon$ is a target error probability, and does so with manageable computational complexity.

\begin{figure}[tbh]
\centerline{\begin{tikzpicture}[
  font=\small, >=stealth', line width=0.75pt,
  infobits0/.style={rectangle, minimum height=6mm, minimum width=20mm, draw=black, fill=gray!10},
  infobits/.style={rectangle, minimum height=6mm, minimum width=12mm, draw=black, fill=gray!10},
  paritybits/.style={rectangle, minimum height=6mm, minimum width=8mm, draw=black, fill=lightgray!50}
]

\node[infobits0] (vb0) at (1,0) {$\wv(0)$};
\node[infobits] (vb1) at (2.6,0) {$\wv(1)$};
\node[paritybits] (vp1) at (3.6,0) {$\pv(1)$};
\node[infobits] (vb2) at (4.6,0) {$\wv(2)$};
\node[paritybits] (vp2) at (5.6,0) {$\pv(2)$};
\node[infobits] (vb3) at (6.6,0) {$\wv(3)$};
\node[paritybits] (vp3) at (7.6,0) {$\pv(3)$};
\draw[|-|] (0,-0.5) to node[midway,below] {$m_0$} (2,-0.5);
\draw[|-|] (2,-0.5) to node[midway,below] {$m_1$} (3.2,-0.5);
\draw[-|] (3.2,-0.5) to node[midway,below] {$l_1$} (4,-0.5);
\draw[|-|] (4,-0.5) to node[midway,below] {$m_2$} (5.2,-0.5);
\draw[-|] (5.2,-0.5) to node[midway,below] {$l_2$} (6,-0.5);
\draw[|-|] (6,-0.5) to node[midway,below] {$m_3$} (7.2,-0.5);
\draw[-|] (7.2,-0.5) to node[midway,below] {$l_3$} (8,-0.5);
\end{tikzpicture}}
\caption{This diagram illustrates the structure of CCS sub-blocks, with their information and parity bits.
Every sub-block is encoded separately before transmission within a slot.}
\label{figure:subblock}
\end{figure}
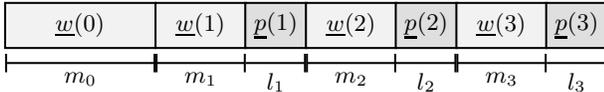
The original CCS scheme works as follows.
An information message is partitioned into several fragments.
Redundancy can be added to every fragment, except for the leading (root) fragment, in the form of parity bits.
These parity bits are formed by taking (random) linear combinations of all the information bits contained in fragments that precede it.
Together, an information fragment and its corresponding parity bits form a sub-block of a size conducive to CS recovery by a commodity solver at the slot level.
The transmission of sub-blocks occurs sequentially, with every slot taking the form of an unsourced MAC problem, albeit one with a much lower dimension.
The structure of sub-blocks appear in Fig.~\ref{figure:subblock}.

Upon completion of slot~$j$, a CS solver is applied to the signal received therein and a list of $\Ka$ sub-blocks is assembled.
Tree decoding is then applied to these lists to stitch fragments into transmitted messages.
As an initial step, the tree decoder selects a root sub-block and compute values for parity bits $\vec{p}(1)$.
Every sub-block in slot~1 that matches these parity bits is attached to the root, thereby producing consistent partial paths.
This process then moves forward.
For every consistent path at stage~$j-1$, parity bits $\vec{p}(j)$ are computed and matching sub-blocks on list~$j$ are attached to this path, forming new branches.
This continues until the last slot is reached.
At this point, every root segment with a unique path to the last slot is deemed a valid tree message; whereas instances where multiple paths from a same root to the last slot survive or all paths from a root halt prematurely are declared decoding failures.

The overall CCS scheme, including both its CS sub-components and the ensuing tree decoding, is described and analyzed in great detail in~\cite{amalladinne2019coded}.
This article also points to a natural tradeoff between error probability and computational complexity, and it offers a principled way to allocate information and parity bits to sub-blocks, so as to achieve good performance.
The treatment presented therein and in follow-up CCS articles~\cite{calderbank2018chirrup,Giuseppe,fengler2019massive,amalladinne2019asynchronous} assumes that decoding takes place in two disjoint stages: sparse recovery followed by tree stitching.
Yet, the structure of CCS invites a more judicious use of information.
Tree decoding can be run in tandem with the CS solver as it progresses through slots.
In particular, the collection of active paths from all the roots to stage~$j-1$ can inform the CS solver at stage~$j$.
This insight and its repercussions are discussed in the next section.

\section{\uppercase{Enhanced Decoding Process}}

As described above, CCS is a divide-and-conquer approach where a large CS problem is broken down into smaller sub-components.
The sparse recovery problem associated with slot~$j$ assumes the following form
\begin{equation} \label{equation:CSInterpretationSlot}
\yv(j) = \mathbf{X}^{(j)} \bv^{(j)} + \zv(j) .
\end{equation}
This equation is analogous to \eqref{syseq}, albeit on a much smaller scale.
We emphasize that $\bv^{(j)}$ remains $\Ka$-sparse, however it is equal to the sum of the message indices corresponding to sub-blocks $\{ \wv_i(j) \pv_i(j) : i \in \mathbf{S}_\mathrm{a} \}$.
The number of columns in $\mathbf{X}^{(j)}$ is $2^{m_j + l_j}$, where $m_j$ and $l_j$ are the numbers of information and parity bits in $\wv(j)$ and $\pv(j)$, respectively.
These parameters are selected to make sure that \eqref{equation:CSInterpretationSlot} is amenable to computationally efficient CS decoders.
This is the way sparse recovery on a slot per slot basis is performed in the original CCS scheme.

In contrast, suppose that sparse recovery and tree stitching are performed concurrently.
Then, by the time the access points is ready to perform sparse recovery on $\yv(j)$, the tree decoder has already identified all the active paths from root fragments to sub-blocks at level~$j-1$.
In addition, it has computed all the possible parity patterns for slot~$j$.
Explicitly, to every active path
\begin{equation*}
\wv(0) \wv(1) \pv(1) \cdots \wv(j-1) \pv(j-1)
\end{equation*}
corresponds a parity pattern $\pv(j)$.
If there are no active paths that lead to a specific parity pattern, then any sub-block at level~$j$ that contains this parity pattern will eventually be discarded by the tree decoder.
It has effectively become an inadmissible pattern based on past observations.

\begin{figure}[tbh]
\centerline{\begin{tikzpicture}
[font=\footnotesize, draw=black, line width=0.75pt,>=stealth',
sub0/.style={rectangle, draw, inner sep=0pt, minimum width=10mm, minimum height=2.5mm},
parity/.style={rectangle, draw, fill=lightgray, inner sep=0pt, minimum size=2.5mm}]

\node (cs1) at (0.00,5.875) {Slot~1};
\node (cs2) at (2.50,5.875) {Slot~2};
\node (cs3) at (5.00,5.875) {Slot~3};

\foreach \v in {0.00,2.50,5.00} {
  \draw[->, line width=1pt]  (\v,4.25) -- (\v,3.875);
  \draw[->, line width=1pt]  (\v,5.625) -- (\v,5.125);
  \draw[dotted, line width=1pt, draw=gray]  (\v-0.25,5.525) -- (\v,5.125);
  \draw[dotted, line width=1pt, draw=gray]  (\v-0.125,5.575) -- (\v,5.125);
  \draw[dotted, line width=1pt, draw=gray]  (\v+0.125,5.575) -- (\v,5.125);
  \draw[dotted, line width=1pt, draw=gray]  (\v+0.25,5.525) -- (\v,5.125);
}

\foreach \v in {0.00} {
  \draw[line width=1pt] (\v-0.5,5) -- (\v-0.625,5) -- (\v-0.625,4.5) -- (\v-0.5,4.5);
  \draw[line width=1pt] (\v+0.375,5) -- (\v+0.5,5) -- (\v+0.5,4.5) -- (\v+0.375,4.5);
  \draw[line width=1pt] (\v+0.575,5) -- (\v+0.575,3.875) -- (\v+0.695,3.875) -- (\v+0.695,5) -- (\v+0.575,5);
}

\foreach \v in {2.50,5.00} {
  \draw[line width=1pt] (\v-0.275,5) -- (\v-0.4,5) -- (\v-0.4,4.5) -- (\v-0.275,4.5);
  \draw[line width=1pt] (\v+0.15,5) -- (\v+0.275,5) -- (\v+0.275,4.5) -- (\v+0.15,4.5);
  \draw[line width=1pt] (\v+0.35,5) -- (\v+0.35,4.325) -- (\v+0.47,4.325) -- (\v+0.47,5) -- (\v+0.35,5);
}

\foreach \p/\c in {3.50/1, 3.125/2, 2.75/3, 2.375/4, 2/5} {
  \node[sub0] (subcs1\c) at (0.0,\p) {};
  \node[sub0] (subcs2\c) at (2.50,\p) {};
  \node[parity] (parity0\c) at (2.875,\p) {};
  \node[sub0] (subcs3\c) at (5.00,\p) {};
  \node[parity] (parity1\c) at (5.125,\p) {};
  \node[parity] (parity2\c) at (5.375,\p) {};
}

\node (list1) at (0.00,1.625) {List~1};
\node (list2) at (2.50,1.625) {List~2};
\node (list3) at (5.00,1.625) {List~3};

\draw [line width=1pt,->] plot[smooth, tension=.5] coordinates {(0.5,1.625) (1.25,2.125) (1.375,4) (1.875,4.75)};
\draw [line width=1pt,->] plot[smooth, tension=.5] coordinates {(3.0,1.625) (3.75,2.125) (3.875,4) (4.375,4.75)};
\draw [line width=1pt,dashed] plot[smooth, tension=.5] coordinates {(5.5,1.625) (6.25,2.125) (6.375,4.25)};

\node[rotate=90] (prune1) at (1.0625,3.25) {column pruning};
\node[rotate=90] (prune2) at (3.5625,3.25) {column pruning};
\node[rotate=90] (prune3) at (6.0625,3.25) {column pruning};
\end{tikzpicture}}
\caption{This notional diagram shows how a tree decoder that runs in parallel with the sequential sparse recovery process can inform the latter about inadmissible parity patterns.
This, in turn, leads to the preemptive pruning of the sensing matrices, which enhances performance and reduces complexity.}
\label{figure:eCCS}
\end{figure}
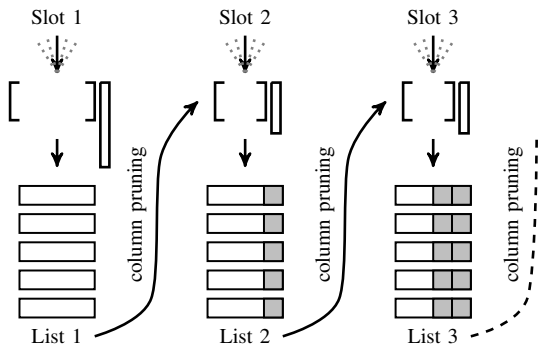

This realization introduces an algorithmic opportunity for performance enhancement.
Instead of waiting for this information to be employed by the tree decoder, it can be used preemptively during the sparse recovery of slot~$j$.
In particular, all the columns in $\mathbf{X}^{(j)}$ that are attached to sub-blocks containing inadmissible parity patterns can be pruned.
Let the set of possible parity patterns at stage~$j$, given past observations, be denoted by $\mathcal{P}_j$.
Then, the number of columns in the dynamically pruned version of $\mathbf{X}^{(j)}$ becomes $2^{m_j} |\mathcal{P}_j|$, rather than the original $2^{m_j + l_j}$.
This conceptual algorithm improvement is illustrated in Fig.~\ref{figure:eCCS}.

\textbf{\textit{Complexity Reduction:}}
It is possible to assess the expected dimensionality reduction delivered via this enhanced decoding algorithm by tracking the expected number of consistent partial paths seen at various stages during the decoding process.
To do so, we leverage the approximate tree code analysis found in~\cite{amalladinne2019coded} under the simplifying assumption that $\wv_i (j) \neq \wv_k (j)$ for any $i \neq k$.
We note that the same article offers an exact (and cumbersome) analysis of this particular problem.
However, the aforementioned assumption is valid with high probability at every stage~$j$ where $m_j$ is large, and the ensuing curves are representative for operating regimes of interest.
The complexity reduction analysis also assumes that the CS lists are error free. 
Under these conditions, the number of active paths from a single root to slot~$j$ is given below.

\begin{proposition}[\hspace{-0.05em}\cite{amalladinne2019coded}] \label{proposition:ExpectedValuesDistinctFragments}
The expected number of erroneous paths that survive stage~$j$, which we denote by $L_j$, is
\begin{equation} \label{expLi}
\mathbb{E} [ L_j ]
= \textstyle \sum_{q=1}^{j} \left( \Ka^{j-q}(\Ka-1) \prod_{\ell=q}^j p_{\ell} \right)
\end{equation}
where $p_{\ell} = 2^{-l_{\ell}}$.
\end{proposition}
Since there are $\Ka$ root fragments, the expected number of consistent partial paths is $P_j = \Ka + \Ka \mathbb{E} [ L_j ]$.
If we further assume that parity patterns are independent from one another and $P_j$ has concentrated around its mean, we get
\begin{equation*}
|\mathcal{P}_j|
\approx 2^{l_j} \left( 1 - ( 1 - 2^{-l_j})^{P_j} \right) .
\end{equation*}
The expected reduction ratio of the number of columns for the sensing matrix at slot~$j$ is then equal to $1 - ( 1 - 2^{-l_j})^{P_j}$.

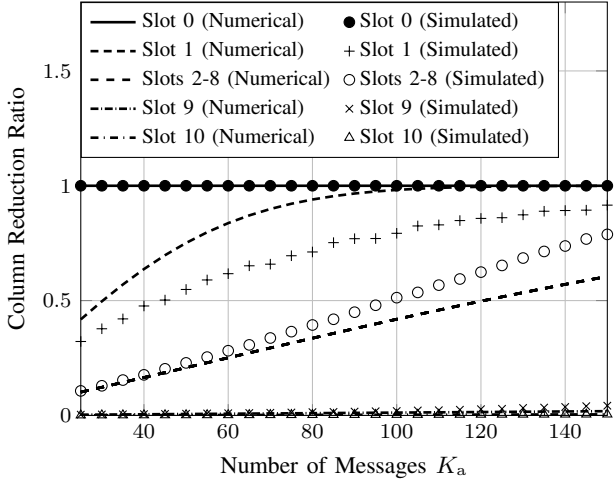
\begin{figure}[tbh]
\centerline{
\begin{tikzpicture}

\begin{axis}[%
font=\footnotesize,
width=7cm,
height=5.5cm,
scale only axis,
xmin=25,
xmax=150,
xlabel={\small Number of Messages $\Ka$},
xmajorgrids,
ymin=0,
ymax=1.8,
ylabel={\small Column Reduction Ratio},
ymajorgrids,
legend columns=2, 
legend style={at={(0,1)},anchor=north west,legend cell align=left,
/tikz/column 2/.style={column sep=3pt},
        },
]

\addplot [color=black,solid,line width=1.0pt]
  coordinates {
(25, 1.00000) (30, 1.00000) (35, 1.00000) (40, 1.00000) (45, 1.00000) (50, 1.00000) (55, 1.00000) (60, 1.00000) (65, 1.00000) (70, 1.00000) (75, 1.00000) (80, 1.00000) (85, 1.00000) (90, 1.00000) (95, 1.00000) (100, 1.00000) (105, 1.00000) (110, 1.00000) (115, 1.00000) (120, 1.00000) (125, 1.00000) (130, 1.00000) (135, 1.00000) (140, 1.00000) (145, 1.00000) (150, 1.00000)
};
\addlegendentry{Slot 0 (Numerical)};

\addplot [only marks,mark=*]
  coordinates {
(25, 1.00000) (30, 1.00000) (35, 1.00000) (40, 1.00000) (45, 1.00000) (50, 1.00000) (55, 1.00000) (60, 1.00000) (65, 1.00000) (70, 1.00000) (75, 1.00000) (80, 1.00000) (85, 1.00000) (90, 1.00000) (95, 1.00000) (100, 1.00000) (105, 1.00000) (110, 1.00000) (115, 1.00000) (120, 1.00000) (125, 1.00000) (130, 1.00000) (135, 1.00000) (140, 1.00000) (145, 1.00000) (150, 1.00000)
};
\addlegendentry{Slot 0 (Simulated)};

\addplot [color=black,densely dashed,line width=1.0pt]
  coordinates {
(25, 0.41804) (30, 0.49668) (35, 0.57002) (40, 0.63716) (45, 0.69757) (50, 0.75100) (55, 0.79749) (60, 0.83732) (65, 0.87092) (70, 0.89882) (75, 0.92167) (80, 0.94010) (85, 0.95475) (90, 0.96624) (95, 0.97511) (100, 0.98188) (105, 0.98697) (110, 0.99075) (115, 0.99351) (120, 0.99550) (125, 0.99692) (130, 0.99792) (135, 0.99861) (140, 0.99908) (145, 0.99940) (150, 0.99961)
};
\addlegendentry{Slot 1 (Numerical)};

\addplot [only marks,mark=+]
  coordinates {
(25,0.3218750) (30,0.3773437) (35,4.203125e-01) (40,4.765625e-01) (45,5.023437e-01) (50,5.484375e-01) (55,5.898438e-01) (60,6.171875e-01) (65,6.515625e-01)
(70,6.585937e-01) (75,6.953125e-01) (80,7.117187e-01) (85,7.523437e-01) (90,7.695313e-01) (95,7.703125e-01)
(100,7.929688e-01) (105,8.250000e-01) (110,8.296875e-01) (115,8.484375e-01) (120,8.578125e-01)(125,8.609e-01)(130,8.734e-01)(135,8.8906e-01)(140,8.9218e-01)(145,8.9375e-01)(150,9.1562e-01)

};
\addlegendentry{Slot 1 (Simulated)};

\addplot [color=black,dashed,line width=1.0pt]
  coordinates {
(25, 0.10149) (30, 0.12253) (35, 0.14374) (40, 0.16507) (45, 0.18649) (50, 0.20797) (55, 0.22947) (60, 0.25095) (65, 0.27240) (70, 0.29377) (75, 0.31504) (80, 0.33617) (85, 0.35715) (90, 0.37794) (95, 0.39851) (100, 0.41885) (105, 0.43893) (110, 0.45873) (115, 0.47823) (120, 0.49741) (125, 0.51626) (130, 0.53476) (135, 0.55289) (140, 0.57064) (145, 0.58800) (150, 0.60497)
};
\addlegendentry{Slots 2-8 (Numerical)};

\addplot [only marks,mark=o]
  coordinates {(25,1.067801e-01) (30,1.290179e-01) (35,1.539342e-01) (40,1.766462e-01) (45,2.029297e-01) (50,2.287109e-01) (55,2.539342e-01) (60,2.817243e-01) (65,3.071429e-01)
(70,3.371094e-01) (75,3.647879e-01) (80,3.937221e-01) (85,4.186663e-01) (90,4.494420e-01) (95,4.797433e-01)
(100,5.127511e-01) (105,5.356027e-01) (110,5.672991e-01) (115,5.939732e-01) (120,6.239955e-01)(125,6.5306e-01)(130,6.8526e-01)(135,7.1395e-01)(140,7.3766e-01)(145,7.6813e-01)(150,7.8825e-01)
};
\addlegendentry{Slots 2-8 (Simulated)};

\addplot [color=black,densely dashdotted,line width=1.0pt]
  coordinates {
(25, 0.00306) (30, 0.00367) (35, 0.00428) (40, 0.00489) (45, 0.00551) (50, 0.00612) (55, 0.00674) (60, 0.00735) (65, 0.00797) (70, 0.00858) (75, 0.00920) (80, 0.00981) (85, 0.01043) (90, 0.01104) (95, 0.01166) (100, 0.01228) (105, 0.01290) (110, 0.01352) (115, 0.01413) (120, 0.01475) (125, 0.01537) (130, 0.01599) (135, 0.01661) (140, 0.01723) (145, 0.01785) (150, 0.01847)
};
\addlegendentry{Slot 9 (Numerical)};

\addplot [only marks,mark=x]
  coordinates {(25,3.332520e-03) (30,4.180908e-03) (35,4.901123e-03) (40,5.511475e-03) (45,6.542969e-03) (50,7.434082e-03) (55,8.605957e-03) (60,9.436035e-03) (65,1.044312e-02)
 (70,1.149292e-02) (75,1.278687e-02) (80,1.430054e-02) (85,1.522217e-02) (90,1.674194e-02) (95,1.811523e-02)
(100,1.921997e-02) (105,2.165527e-02) (110,2.257080e-02) (115,2.467041e-02) (120,2.714844e-02)(125,2.8845e-02)(130,3.2532e-02)(135,3.4167e-02)(140,3.6804e-02)(145,4.0258e-02)(150,4.2333e-02)
};
\addlegendentry{Slot 9 (Simulated)};

\addplot [color=black,dashdotted,line width=1.0pt]
  coordinates {
(25, 0.00076) (30, 0.00092) (35, 0.00107) (40, 0.00122) (45, 0.00137) (50, 0.00153) (55, 0.00168) (60, 0.00183) (65, 0.00199) (70, 0.00214) (75, 0.00229) (80, 0.00244) (85, 0.00260) (90, 0.00275) (95, 0.00290) (100, 0.00306) (105, 0.00321) (110, 0.00336) (115, 0.00352) (120, 0.00367) (125, 0.00382) (130, 0.00398) (135, 0.00413) (140, 0.00428) (145, 0.00443) (150, 0.00459)
};
\addlegendentry{Slot 10 (Numerical)};

\addplot [only marks,mark=triangle]
  coordinates {(25,7.644653e-04) (30,9.170532e-04) (35,1.072693e-03) (40,1.228333e-03) (45,1.380920e-03) (50,1.538086e-03) (55,1.689148e-03) (60,1.841736e-03) (65,1.997375e-03)
 (70,2.151489e-03) (75,2.307129e-03) (80,2.476501e-03) (85,2.627563e-03) (90,2.786255e-03) (95,2.940369e-03)
(100,3.100586e-03) (105,3.288269e-03) (110,3.399658e-03) (115,3.573608e-03) (120,3.771973e-03)(125,3.9337e-03)(130,4.0740e-3)(135,4.2205e-03)(140,4.3854e-03)(145,4.5380e-03)(150,4.7241e-03)
};
\addlegendentry{Slot 10 (Simulated)};

\addplot [color=black,dashed,line width=1.0pt]
  coordinates {
(25, 0.10149) (30, 0.12253) (35, 0.14374) (40, 0.16507) (45, 0.18649) (50, 0.20797) (55, 0.22947) (60, 0.25095) (65, 0.27240) (70, 0.29377) (75, 0.31504) (80, 0.33617) (85, 0.35715) (90, 0.37794) (95, 0.39851) (100, 0.41885) (105, 0.43893) (110, 0.45873) (115, 0.47823) (120, 0.49741) (125, 0.51626) (130, 0.53476) (135, 0.55289) (140, 0.57064) (145, 0.58800) (150, 0.60497)
};

\addplot [color=black,dashed,line width=1.0pt]
  coordinates {
(25, 0.10149) (30, 0.12253) (35, 0.14374) (40, 0.16507) (45, 0.18649) (50, 0.20797) (55, 0.22947) (60, 0.25095) (65, 0.27240) (70, 0.29377) (75, 0.31504) (80, 0.33617) (85, 0.35715) (90, 0.37794) (95, 0.39851) (100, 0.41885) (105, 0.43893) (110, 0.45873) (115, 0.47823) (120, 0.49741) (125, 0.51626) (130, 0.53476) (135, 0.55289) (140, 0.57064) (145, 0.58800) (150, 0.60497)
};

\addplot [color=black,dashed,line width=1.0pt]
  coordinates {
(25, 0.10149) (30, 0.12253) (35, 0.14374) (40, 0.16507) (45, 0.18649) (50, 0.20797) (55, 0.22947) (60, 0.25095) (65, 0.27240) (70, 0.29377) (75, 0.31504) (80, 0.33617) (85, 0.35715) (90, 0.37794) (95, 0.39851) (100, 0.41885) (105, 0.43893) (110, 0.45873) (115, 0.47823) (120, 0.49741) (125, 0.51626) (130, 0.53476) (135, 0.55289) (140, 0.57064) (145, 0.58800) (150, 0.60497)
};

\addplot [color=black,dashed,line width=1.0pt]
  coordinates {
(25, 0.10149) (30, 0.12253) (35, 0.14374) (40, 0.16507) (45, 0.18649) (50, 0.20797) (55, 0.22947) (60, 0.25095) (65, 0.27240) (70, 0.29377) (75, 0.31504) (80, 0.33617) (85, 0.35715) (90, 0.37794) (95, 0.39851) (100, 0.41885) (105, 0.43893) (110, 0.45873) (115, 0.47823) (120, 0.49741) (125, 0.51626) (130, 0.53476) (135, 0.55289) (140, 0.57064) (145, 0.58800) (150, 0.60497)
};

\addplot [color=black,dashed,line width=1.0pt]
  coordinates {
(25, 0.10149) (30, 0.12253) (35, 0.14374) (40, 0.16507) (45, 0.18649) (50, 0.20797) (55, 0.22947) (60, 0.25095) (65, 0.27240) (70, 0.29377) (75, 0.31504) (80, 0.33617) (85, 0.35715) (90, 0.37794) (95, 0.39851) (100, 0.41885) (105, 0.43893) (110, 0.45873) (115, 0.47823) (120, 0.49741) (125, 0.51626) (130, 0.53476) (135, 0.55289) (140, 0.57064) (145, 0.58800) (150, 0.60497)
};

\addplot [color=black,dashed,line width=1.0pt]
  coordinates {
(25, 0.10149) (30, 0.12253) (35, 0.14374) (40, 0.16507) (45, 0.18649) (50, 0.20797) (55, 0.22947) (60, 0.25095) (65, 0.27240) (70, 0.29377) (75, 0.31504) (80, 0.33617) (85, 0.35715) (90, 0.37794) (95, 0.39851) (100, 0.41885) (105, 0.43893) (110, 0.45873) (115, 0.47823) (120, 0.49741) (125, 0.51626) (130, 0.53476) (135, 0.55289) (140, 0.57064) (145, 0.58800) (150, 0.60497)
};
\end{axis}

\end{tikzpicture}}
\caption{This graph illustrates the drastic reduction associated with matrix pruning in the enhanced decoding process for the parity allocation in \eqref{equation:ParityAllocation}.
For every slot, the curve reflects the (approximate) number of columns in the pruned sensing matrix over the original width of the matrix.
The reduction is much more pronounced for later stages.}
\label{figiure:ColumnReduction}
\end{figure}

To further demonstrate the benefits of pruning technique, we consider one of the optimized parity allocation sequence given in~\cite{amalladinne2019coded},
\begin{equation} \label{equation:ParityAllocation}
(l_1, l_2, \ldots, l_{10}) = (6, 8, 8, 8, 8, 8, 8, 8, 13, 15) .
\end{equation}
Figure~\ref{figiure:ColumnReduction} shows a significant reduction in the size of pruned matrices after the first few stages.
Similar results are observed for alternate parity allocations.
This behavior directly translates into a complexity reduction for the CS solvers, especially at the later stages.

\textbf{\textit{Additional Implications:}}
The dynamic pruning of the sensing matrices has implications beyond the matrix width reduction described above.
First, we stress once again that the analysis presented above naively assumes that the CS output lists contain all valid segments.
This may not always be the case.
The dynamic pruning seems to affect the slot CS decoding in, at least, three different ways.
\begin{enumerate}
\item When the previous stages have identified all the correct sub-blocks, the sensing matrix for the current stage is trimmed down in a way that is consistent with the problem statement.
This reduces the search space for the CS solver and improves its performance.
\item If an erroneous partial path survives until stage~$j-1$, then the pruned sensing matrix at stage~$j$ retains all the columns with parity patterns that are consistent with this erroneous path, but discards other columns.
This steers the CS solver towards a list that is more likely to include sub-blocks that are consistent with the erroneous path.
This increases the propensity for error propagation, with erroneous paths staying alive longer on average.
\item If a valid sub-block is omitted from a CS list, then the corresponding parity pattern may disappear.
When this is the case, the received vector for the subsequent slot is no longer of the form $\yv(j) = \mathbf{X}^{(j)} \bv^{(j)} + \zv(j)$ because of the missing columns.
This results in noise amplification for the other messages being decoded.
\end{enumerate}
Despite some negative aspects of the enhanced decoding process for CCS, the proposed approach improves overall performance beyond the obvious complexity reduction.
This is illustrated in the next section.

\section{\uppercase{Performance Evaluation}}
\label{section:PerformanceEvaluation}

The simulation results contained in this section adopt a set of parameters that has become widespread on articles related to the unsourced MAC.
While the algorithmic enhancement described above is general, this choice of parameters is conducive to a rapid and fair comparison with alternate schemes.
We examine a system where $\Ka \in [10 : 300]$ and $B = 75$~bits.
The total number of channel uses is 22,517.
The message recovery task is partitioned into 11~stages.
Performance is reported in the form of the minimum $E_{\mathrm{b}}/N_0$ required to achieve per user error probability of $P_{\mathrm{e}} = 0.05$.

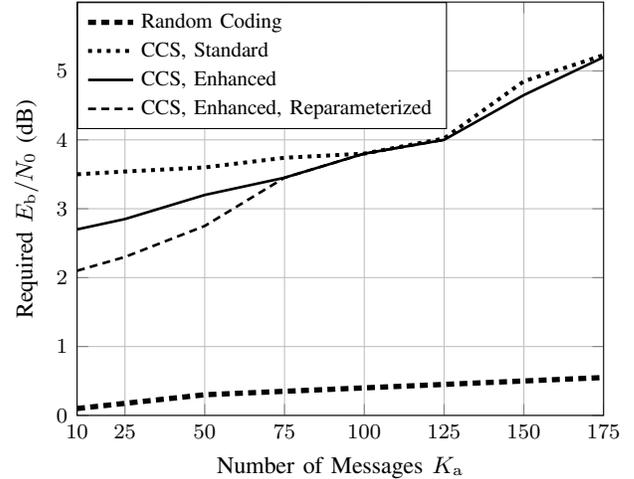
\begin{figure}[tbh]
\centerline{
\begin{tikzpicture}

\begin{axis}[%
font=\footnotesize,
width=7cm,
height=5.5cm,
scale only axis,
xmin=10,
xmax=175,
xtick = {10,25,50,75,...,175},
xlabel={\small Number of Messages $\Ka$},
xmajorgrids,
ymin=0,
ymax=6,
ytick = {0,1,2,...,5},
ylabel={\small Required $E_{\mathrm{b}}/N_0$ (dB)},
ymajorgrids,
legend style={at={(0,1)},anchor=north west,draw=black,fill=white,legend cell align=left}
]

\addplot [color=black,densely dashed,line width=2.0pt]
  table[row sep=crcr]{
  10 0.1
 25	0.25\\
50	0.3\\
75	0.35\\
100	0.4\\
125	0.45\\
150	0.5\\
175	0.55\\
};
\addlegendentry{Random Coding};

\addplot [color=black,dotted,line width=1.5pt]
  table[row sep=crcr]{
 10 3.5\\ 
25	3.54\\
50	3.6\\
75	3.74\\
100	3.8\\
125	4.02\\
150	4.85\\
175	5.23\\
};
\addlegendentry{CCS, Standard};

\addplot [color=black,solid,line width=1.0pt]
  table[row sep=crcr]{
10  2.7\\  
25	2.85\\
50	3.2\\
75	3.45\\
100 3.8\\
125 4\\
150 4.65\\
175 5.2\\
};
\addlegendentry{CCS, Enhanced};

\addplot [color=black,densely dashed,line width=1.0pt]
  table[row sep=crcr]{
10  2.1\\  
25 2.3\\
50 2.75\\
75	3.45\\
100 3.8\\
125 4\\
};
\addlegendentry{CCS, Enhanced, Reparameterized};

\end{axis}

\end{tikzpicture}%
}
\caption{The enhanced decoding algorithm for CCS yields better per user probability of error and reduces complexity.
Additional gains are possible through reparameterization.}
\label{figure:PerformanceComparisonCCS}
\end{figure}
In previous articles on CCS, the number of channel uses is partitioned equally into 11 slots, with each slot having length 2047. 
The columns of the base $\mathbf{X}^{(j)}$ (before pruning) are judiciously selected codewords from the (2047,23) BCH codebook~\cite{amalladinne2019coded}.
Note that these binary codewords are centered and renormalized to become proper signals. 
The first two curves on Fig.~\ref{figure:PerformanceComparisonCCS} correspond to the original scheme first reported in~\cite{amalladinne2018couple} and the performance improvement associated with the enhanced algorithm introduced above.
For these curves, the length of each coded sub-block is set to 14 for $\Ka \in [10:125]$ and 15 for $\Ka > 125$.
In both cases, we allow one extended iteration whereby the strongest messages are removed from the received signal and the decoding algorithm is performed a second time on the residual signal.

The enhanced decoding introduces new possibilities in terms of system design.
The size of the sub-blocks in CCS is constrained by the width of the sensing matrix $2^{15}$, with the understanding that this is close to the limit of what a commodity CS solver can handle on a conventional computer.
However, under the dynamic pruning of the sensing matrices, these design parameters can be revisited.
For instance, one could devote more channel uses to early slots where the sampling matrices remain essentially untouched, with the later stages necessarily receiving fewer symbols.
Alternatively, the allocation of information and parity bits per slot can be re-optimized, taking into consideration the eventual dimensionality reduction produced by the pruning process.
Due to space restrictions, we cannot discuss these possibilities at length.
Still, we include a third curve on Fig.~\ref{figure:PerformanceComparisonCCS} to showcase how a reparameterization of the system leads to improvements.

\section{\uppercase{Conclusions and Future Work}}

This article highlights an algorithmic improvement to the decoding process for CCS based on the structure of the encoding process.
This improvements leads to both a decrease in the per user probability of error and a significant reduction in computational complexity.
Under this technique, additional gains can be obtained by further optimizing system parameters.
It appears that the same information structure can be leveraged in alternate versions of CCS, including those relying on AMP.

\bibliographystyle{IEEEbib}
\bibliography{IEEEabrv,MACcollision}

\end{document}